# Influence of Relic Neutrinos on Beta Radioactivity

A.G.Parkhomov
Institute for Time Nature Explorations.
Lomonosov Moscow State University, Moscow, Russia.
www.chronos.msu.ru

Results of calculations of distribution and motion of dark matter are presented. Considering neutrino as one of the components of dark matter and taking into account peculiar features of the interactions of slow neutrinos with matter, allow to make the conclusion that they may have tangible manifestations not only in the depths of the Universe but also on the Earth. Experimental results confirming predicted effects are described, including periodic variations of the β radioactivity as well as count rate bursts for a β radioactive source placed at the focal point of a parabolic mirror. Based on the data of astronomical observations, estimates of the mass of the particles influencing on β radioactivity (about 20 $eV$) and their flux density (about $10^{13}$ particles/cm$^2$ s) have been made. The discrepancy between our mass estimate and the 2 $eV$ limit for the neutrino mass, established in the tritium experiments, is discussed.

Dark matter consists of a substance characterized by a diffused state and very weak interactions with matter, which makes its detection extremely difficult [1]. It manifests itself as gravitational influence caused by a mass, hidden from direct observations, on motion of objects accessible to astronomical observations. Particles of dark matter should possess non-zero mass since the ability to concentrate in the gravitational fields of galaxies and stars implies objects with speeds that do not exceed one thousandth of the speed of light.

Dark matter has multiple components and its study is limited largely to the realm of hypotheses. At present, the prevalent point of view is that the main constituent parts of dark matter are hypothetical particles, which were predicted by the "Grand Unification Theory" and quantum chromodynamics. No experimental evidence of the existence of these particles is available. Of all the possible components of dark matter only neutrinos which were produced at the very early stages of the existence of the Universe ("relic neutrinos"), are to some extent a familiar territory for modern science [1,2,23]. Energy of these neutrinos is so small that even if they possess tiny mass their speed will be much less than the speed of light. One cannot exclude that there are other mechanisms leading to the appearance in the Universe of neutrinos with very small energies. Because of this, it seems expedient to give such objects common name "slow neutrinos", which emphasizes their most important difference from relativistic neutrinos arising in β decays and nuclear reactions. It was estimated that neutrinos constitute about 1% of all dark matter. Even this should be considered a significant figure. As an illustration, cosmologists estimate the number of relic neutrinos to exceed the number of atoms in the Universe by a factor of about billion.

Despite current state of knowledge about dark matter, available astronomical data is sufficient to calculate its distribution and movement. Considering neutrino as one of the components of dark matter and taking into account peculiar features of the interactions of slow neutrinos with matter allow to reach conclusion that they may have tangible manifestations not only in the depths of the Universe but also on the Earth. Experimental results confirming predicted effects were obtained.

### Distribution and motion of dark matter

Dark matter in gravitationally bound systems can be in a state, which is "spread" through space, only if it moves with speeds in some well defined range. Since at present there are no reasons to question universal applicability of the laws of gravitation, it seems reasonable to assume that trajectories of constituent objects of dark matter, regardless of their nature, do not differ from trajectories of other celestial bodies such as stars, planets, asteroids, cosmic dust etc. The orbits of these particles can be calculated by means of traditional methods of celestial mechanics, taking into account quantum effects in the case of high concentrations of particles. Such calculations can be found in [3, 4], together with the following conclusions.

Objects of dark matter are members of the same hierarchy of gravitationally bound systems which also contains observable objects: systems of the Earth, the Sun, star clusters, galaxies, clusters of galaxies. The surface of the Earth is reached by fluxes of dark matter belonging to all of these systems. Each of them has its inherent range of particle speeds. Near the surface of our planet these speeds fill several ranges from 7.9 to several thousands of *km/s* (Fig. 1). Angular distributions corresponding to each range are different. Extragalactic particles seem to come from all directions with approximately the same probability (except for a small area near the Sun). The most intensive fluxes of galactic particles come from two areas in particle constellations Aquila and Perseus, as well as from the area surrounding the Sun. Particles orbiting the Sun come mainly from the areas that trail the Sun along the ecliptic. Fluxes of particles orbiting in the Earth-Moon system depend not only on the current configuration of the Earth, Moon and the Sun, but also on their scattering in the Earth. The latter fact gives rise to the complex spatial and temporal patterns of the intensity of these fluxes.

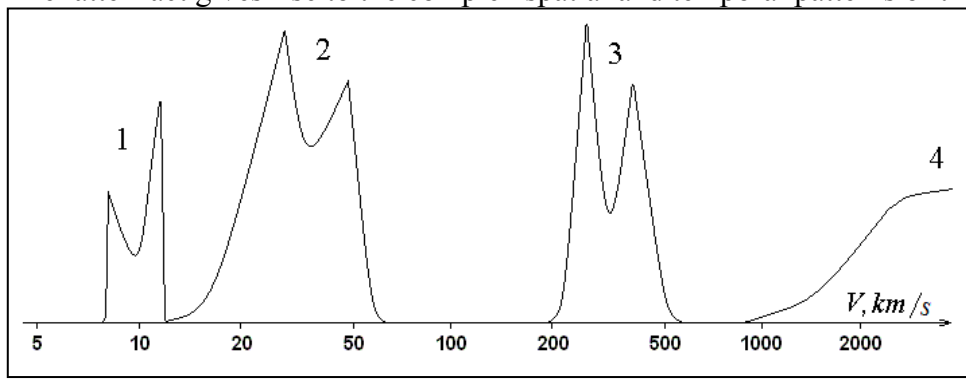

**Fig. 1.** Aspect of distribution of dark matter fluence versus velocities of meeting with a surface of the Earth [3]. 1 – near-Earth particles, 2 - Solar system particles, 3 - galactic, 4 - extragalactic particles.

Fluxes associated with orbital motion are modified by phenomena caused by gravitational focusing of dark matter fluxes when they happen to pass near celestial bodies. This effect is analogous to the well known gravitational lensing of the light. But due to the fact that speeds of particles of dark matter are much smaller than the speed of light, the gravitational fields of celestial bodies affect their trajectories in much stronger manner. The essence of gravitational focusing consists in bending of the paths of the particles, flying past a massive body, towards the line connecting the center of the body and the observer. The extent of the bending action depends on the distance between the trajectory and the center of gravitational force. There exists such a value of that distance when the curved trajectory hits the observer. All the particles passing at this distance from the focusing body "collide" at the point of observation and, as a result, the flux density increases sharply. Gravitational focusing may even lead to concentration of fluxes with "fuzzy" angular distribution, giving rise to almost unidirectional jets with high flux density. Such a phenomenon does not change total number of particles but rather redistributes their trajectories in such a way as to increase the flux density for angles corresponding to the central area, at the expense of the particle density in peripheral areas. Therefore, detection of the focused fluxes requires instruments (similar to telescopes) with sufficiently high angular resolution, because averaging over an area covering the zone of high flux density along with the surrounding zone of low density reduces significantly the magnitude of the effect. Since gravitational focusing takes place only for specific spatial arrangements of the focusing body and the target, which keep moving, the effect should manifest itself as bursts of varying duration.

Results of calculations of amplification ratios for fluxes of galactic particles by typical cosmic objects for a telescope with an angular resolution of $1''$ are presented in Table 1. It is interesting that the "brightest", the most efficiently focusing objects are those which cause most difficulties for their detection by optical means: neutron stars, white dwarfs, black holes. The brightest planet Venus is a rather poor amplifier, but distant planet Neptune has "brightness" superior to that of the Sun; white dwarf Sirius B, invisible to the naked eye, in fluxes of galactic particles outshines completely Sirius A, the brightest star of the optical range.

| Sun | normal star | $3,3 \cdot 10^4$ |
|---|---|---|
| Jupiter | planet | $1,6 \cdot 10^4$ |
| Saturn | planet | $1,2 \cdot 10^4$ |
| Neptune | planet | $1,0 \cdot 10^5$ |
| Venus | planet | 6 |
| Betelgeuse | super-giant star | $2,4 \cdot 10^7$ |
| Sirius A | normal star | $5,6 \cdot 10^{10}$ |
| Sirius B | white dwarf | $4,0.10^{11}$ |
| NP 0531 | neutron star | $1,6 \cdot 10^{11}$ |
| Cygnus X1 | black hole | $4,0 \cdot 10^{11}$ |
| M5 | spherical star cluster | $1,6 \cdot 10^{11}$ |

**Table 1.** Amplification ratios for gravitational focusing of an isotropic flux of particles by some cosmic objects. The speed of particles is taken to be 300 km/s [3].

Gravitational focusing has another important property. If the focusing body (e.g. a star) is stationary relative to the observer, the latter perceives the flux as coming from a ring with an angular diameter of the order of ten arc seconds (in the case of normal stars). If the focusing body moves in the direction orthogonal to the line connecting the body and the observer, the flux appears as two arcs (turning into two points at large enough speeds). Note, that one of the arcs (or points) is seen near the place where the star was located at the time when the flux was passing it by, i.e. many centuries ago. The other point is seen near the place where the star is located at the precise moment when the particles focused by it are reaching the observer, i.e. at the location where the star would be seen if the light propagated with infinite velocity [5].

It seems appropriate to mention here some remarkable results obtained by N.A. Kozyrev. He observed stars by means of a telescope-reflector. A special sensor, non-sensitive to light, was mounted at the focus of the telescope [6]. Kozyrev claimed, that such a telescope allowed to see stars in three positions, one of which corresponding to a direction where they were at the instant of the observation, and not at the moment when they had emitted the light. Based on this, a conclusion was drawn that a signal was being registered which was coming from the star with no propagation delay. However, the reported result can be explained in other way, one which does not undermine the established physical principles, if one assumes that Kozyrev registered dark matter flux focused by gravitational field of moving stars [5].

**Properties of neutrinos as constituent part of dark matter.**

Of all the elementary particles that are "mastered" by modern science, only neutrinos can be cast into the role of one of the components of dark matter. Could neutrino be a particle able to exert perceptible influence on radioactivity or alter course of some other processes? At first sight, this idea seems rather absurd, as neutrinos are reputed to be notoriously hard to catch. However, this feature is characteristic of neutrinos with energy of the order of $1 \, MeV$ and higher, which are produced in β decays and nuclear reactions. Dark matter neutrinos have energies ten orders of magnitude lower and move with speeds that are much slower than the speed of light. Therefore, properties discovered by studying high energy neutrinos should not be transferred unchanged to dark matter neutrinos. It would be similar to the situation in which, for instance, properties of gamma-rays were ascribed to radio-waves, or behavior of alpha particles were considered as a good guide to the dynamics of liquid helium.

The principal difference lies in the fact that flux of slow neutrinos manifests pronounced wave properties. The range of wavelengths overlaps with macroscopic scales and reaches several millimeters. This can be inferred from the estimates of de Broigle wavelengths [3]. Such neutrinos interact not with single nuclei or electrons but rather with their bulk quantities, which leads to a sharp increase in the interaction strength [7, 8, 22]. Behavior of such neutrino in matter is similar to that of light in a transparent medium: processes of refraction, reflection and scattering on impurities can take place, which involve transfer of momentum but virtually no exchange of energy. If surface separating two media is sufficiently smooth, refraction and reflection obey laws of geometrical optics, the fact that hints at possibility of focusing by means of lenses and mirrors. Another way to achieve focusing is by means of the phenomenon of interference.

So, flux density of dark matter neutrinos may increase abruptly during bursts associated with gravitational focusing and, besides, it can be boosted even further using lenses, mirrors and devices based on diffraction. However, is all this enough for neutrinos, even during the bursts, to manifest their existence in a tangible way? The most reliable method of making sure that neutrinos rather than some other particles are being detected, employs nuclear reactions of reverse β decay

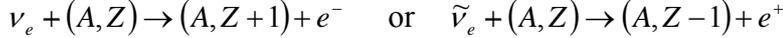

$$\nu_e + (A, Z) \rightarrow (A, Z+1) + e^- \quad \text{or} \quad \widetilde{\nu}_e + (A, Z) \rightarrow (A, Z-1) + e^+$$

This technique is applied to registration of electron neutrinos and antineutrinos with energies about 1 *MeV* and higher [9]. Reactions of reverse β decay with stable nuclides inevitably have energy threshold. It is evident that neutrinos (antineutrinos) with vanishing kinetic energy and very small mass cannot participate in a reverse β decay reaction with stable nuclides. In such a situation, it is necessary to use as a target β radioactive nuclides with no threshold energy. Fluxes of neutrinos with very low energy act on β radioactive nuclei in such a way that in addition to continuous spectrum of spontaneously emitted β particles there appear monoenergetic electrons with an energy exceeding the maximal energy of β spectrum by an amount equal to the neutrino mass (Fig. 2). By using a detector with sufficiently good energy resolution and an amplitude discriminator, these electrons can be registered separately from β particles of spontaneous decay process.

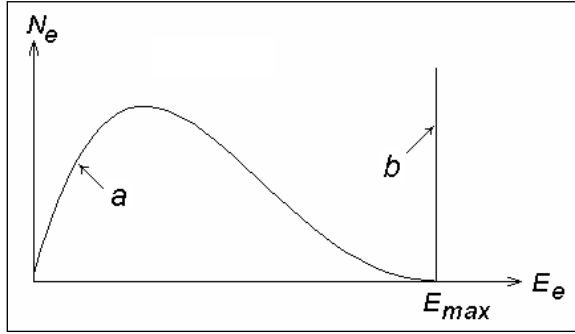

**Fig. 2.** Electron (positron) spectrums of normal (a) and reverse (b) β decays initiated by low energy neutrino. $N_e$ - number of emitted electrons (positrons), $E_e$ - energy of electrons (positrons), $E_{max}$ - maximal energy of β spectrum.

Traditional perception of neutrino as a "phantom" particle strongly suggests that any practical implementation of this method of registration of dark matter neutrinos should be unfeasible. Nevertheless, theoretical considerations in [3] show that the effects caused by cosmic fluxes of slow neutrinos may be quite noticeable. Even a simple application of the same techniques which predict vanishing cross sections for matter interactions of neutrinos with "nuclear" energies, reveals that manifestations of fluxes of slow dark matter neutrinos may be noticeable.

An approach developed by Bethe and Peierls [10] for determination of probability of reverse β decay reactions initiated by neutrino, allowed to obtain for the first time a correct estimate of the cross section of this nuclear reaction. Such an approach, which relies on the assumption of equal probabilities of direct and reverse processes, leads to the expression

$$\sigma = \lambda^3 / Tv, \tag{1}$$

where σ- reaction cross section, $\lambda$ - de Broigle wavelength of neutrinos, $T$- average lifetime of radioactive nuclei, $v$ - speed of neutrinos.

In the case of relativistic neutrinos which are dealt with in nuclear physics, $\lambda = hc/E$ (*h* - Plank constant, *c* - speed of light, *E*– neutrino energy. (1) turns into

$$\sigma = h^3 c^2 / E^3 T. \tag{2}$$

Substituting in (2) typical for nuclear physics values of *E*=1 *MeV*, *T* = 1000 *s*, yields σ= 6·10⁻⁴³ *sm²*, which was confirmed by subsequent experiments.

In the case of neutrinos with very small energies *v<<c*, $\lambda = h/mv$ (*m* - neutrino mass), (1) becomes

$$\sigma = h^3 / m^3 v^4 T. \tag{3}$$

It follows from (3) that

$$n = N\phi\sigma = A\phi h^3/m^3v^4 \,, \tag{4}$$

where $n$ - number of reverse β decays in a second, $A = N/T$ - number of normal β decays in a second (activity of the β source), $N$ - total number of radioactive nuclei, $\phi$ - density of neutrino flux.

To find ratio of the rate of reverse β decays to the rate of spontaneous β radioactivity $K=n/A$, one can make use of (4):

$$K = \phi h^3/m^3v^4. \tag{5}$$

It follows from this formula that, taking neutrino mass to be $m=1\ eV$ ($1.8 \cdot 10^{-36}kg$) and their speed $v=3 \cdot 10^5\ m/s$, an activity increase of 1% requires a density flux $\phi = 6 \cdot 10^{13}\ m^{-2}s^{-1}$. An analysis of available experimental data allows to make an alternative estimate of the neutrino mass, which proves to be different from the accepted value (see below).

An important feature of relation (5) is its independence from half-lives of the nuclei. Any β radioactive sources placed into fluxes of slow neutrinos of equal intensity will experience the same relative increase in their decay rates.

The above analysis of properties of slow neutrinos implies that when the Earth passes through regions with varying speeds or densities of neutrino flux, different β radioactive sources should exhibit relative changes in their activity of equal magnitude.

These calculations are not supposed to be accurate or strict, but rather are intended to show in a clear way that galactic fluxes of neutrinos represent a tangible cosmic factor. An estimate of the extent to which the solar and near-Earth neutrino fluxes affect radioactive decays, does not seem to be feasible at the moment.

### Experimental confirmations

Until recently, it was generally accepted that nuclear decays (except for e-captures) are caused exclusively by intranuclear processes, whose course can not be noticeably altered by common external factors (electromagnetic, thermal, acoustic etc). Therefore, any measurements of the rates of radioactive decays should reveal only an exponential decrease along with superimposed chaotic fluctuations obeying Poisson distribution. Lately however, after it became possible to conduct accurate long-term measurements, some experimental results, which appear to indicate the existence of periodic and sporadic deviations from the exponential law, were obtained. A number of independent researchers [3, 11-16] reported that activity of different sources ($^{60}Co,\ ^{32}Si,\ ^{90}Sr\text{-}^{90}Y,$ decay products of $^{226}Ra$) varies with a period of 1 *year* and a magnitude about 0.3% of the average value (Fig. 3). Moreover, sudden bursts in the activity of $^{60}Co$ and $^{40}K$ samples placed at the focal point of a parabolic mirror were detected [3, 12, 17] (Fig. 4).

Note that periodic variations and bursts were observed only in experiments with β radioactive sources and no such phenomena were reported for α radioactivity. This fact indicates existence of a link between such effects and neutrino fluxes, because neutrinos (antineutrinos) are involved in β processes and do not take part in α decays. In [13], it was suggested that oscillations of β decay rates are associated with the changes of the flux density of the neutrinos emitted by the Sun as a result of the variations of the distance between the Earth and the Sun, caused by the orbital motion of the Earth. But this idea looks rather dubious due to the extremely weak interactions with matter of solar neutrinos characterized by energies of 1 *MeV* and higher. For example, in the experiment of Davies [18] a 610 ton detector registered about 1 solar neutrino daily.

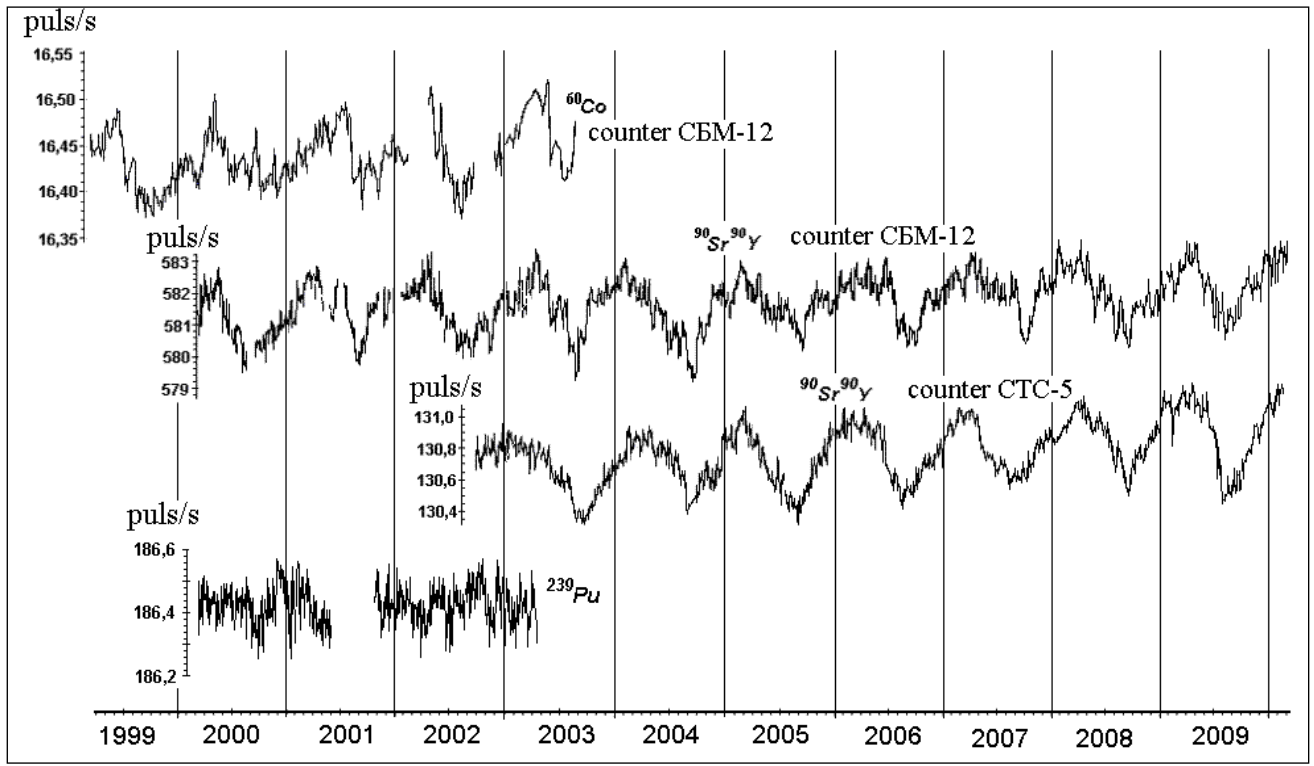

**Fig. 3.** Count rates for $^{60}Co$ and $^{90}Sr$-$^{90}Y$ sources as measured by G-M counters, adjusted for decrease in the radioactivity with half-lives of 5.26 and 27.7 years, as well as count rate for a $^{239}Pu$ source, as measured by a silicon detector [3, 16]. Curves corresponding to β sources exhibit an easy to see annual rhythm, whereas measurements of alpha decays reveal no periodic variations.

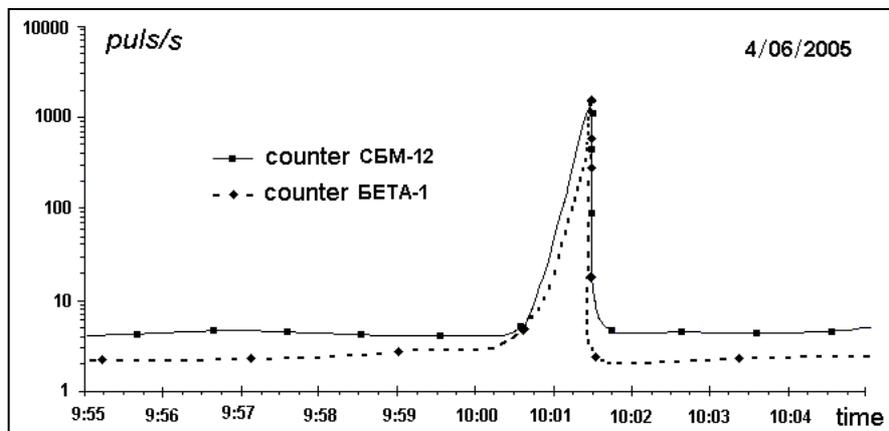

**Fig. 4.** An example of a burst of the count rate of a β source mounted at the focal point of a mirror with parabolic surface. Radiation from a $^{60}Co$ sample was registered by means of two independent Geiger counters [3, 17].

As was already mentioned, besides neutrinos emitted in nuclear reactions, outer space contains a big number of neutrinos with extremely low energies. Theoretical estimate, presented above, of the influence of such neutrinos flux on β radioactivity shows that it may prove quite noticeable. The independence of (5) from half-lives of radionuclides and uniformity of magnitudes of the experimentally found annual variations of the activity for all radionuclides, which were tested for the presence of this effect, can be regarded as another bit of evidence in favor of the hypothesis that it represents manifestation of cosmic fluxes of slow neutrinos.

Combining results of astronomical observations with expressions (4) and (5) enables to estimate mass of neutrino and density of their flux. From (4) it can be seen that magnitude of the effect depends strongly on the value of speed. Let us suppose that the main reason for the existence of annual variations of the activity can be formulated as a basic fact, that the speed of a neutrino flux, arriving at the Solar system from outer space, measured relative to the Earth is a sum of its speed relative to the Sun and the speed of the orbital motion of the Earth around the Sun. According to [4], dark matter flux in the region of the Solar system, including neutrinos, has velocity about $3 \cdot 10^5$ *m/s* and its direction is mostly orthogonal to the direction of the Sun's motion in the Galaxy, characterized by velocity of $2.5 \cdot 10^5$ $m/s$. The Earth circles around the Sun at $3 \cdot 10^4$ $m/s$. Making use of all these figures, it is a straightforward exercise to determine that the speed of the Earth relative to a typical flux of galactic neutrinos changes in a given year from $3.7 \cdot 10^5$ to $4.1 \cdot 10^5$ $m/s$. This range of speeds corresponds to variations in the activity $A$ with magnitude of $0.006A$. The ratio of the activity variations to the velocity change is $\Delta n/\Delta v = -1{,}5 \cdot 10^{-7}$ $A$. On the other hand, differentiating (4) yields $\Delta n/\Delta v = -4A\phi h^3/m^3 v^5$. Taking into account that $\phi = \rho v/m$, where $\rho$ is mass density of neutrinos in the near-Earth space, leads to

$$m = 9{,}4 \cdot 10^{-24} \rho^{1/4}/v \qquad (6)$$

In the region of space occupied by the Solar system, the density of galactic dark matter is about $5 \cdot 10^{-22}$ $kg /m^3$ [3, 4]. If neutrinos account for 1% then $\rho = 5 \cdot 10^{-24}$ $kg/m^3$, and for velocity value of $4 \cdot 10^5$ $m/s$, $m = 3.5 \cdot 10^{-35}$ $kg$ (20 $eV$) and flux density $\phi = 5.7 \cdot 10^{16} m^{-2} s^{-1}$. The ratio of the number of β decays induced by the neutrino flux to the number of spontaneous decays $K = 0.015$.

Varying the assumed neutrinos' share in the total mass of dark matter from 0.1 to 10% produces neutrino mass estimates in the range 11- 35 $eV$.

The neutrino mass estimate obtained in such a way is consistent with results of the experiments involving diffraction gratings, which allowed to determine wavelengths of the agent that affected β radioactive samples. Several intervals with wavelengths of 5.2...7.3; 46...68; 0.3...0.5 $mm$; 1.4...2.0 $mm$ were identified [3]. Assuming that maximal wavelength of 2 $mm$ corresponds to the lowest possible orbital speed of 7.9 $km/s$, the de Broigle formula $\lambda = h/mv$ gives an estimate for the mass of diffracted particles $m = 23$ $eV$. For such a value of the mass, the above wavelength intervals correspond to speeds ranges of 7.9-11.2; 31-53; 230-350; 2100-3000 $km/s$, which agree with the predicted distribution of the speeds of particles, moving along bound orbits in the gravitational fields of the Earth, the Sun and the Galaxy, and coming from intergalactic space (fig. 1). It indicates that the registered agent really is a component of dark matter.

The assumption about the possibility of focusing of fluxes of the neutrino component of dark matter found its confirmation in observations of bursts of β radioactivity of a source located in the focus of a parabolic mirror (fig. 4). Focusing is feasible if the agent is capable of mirror reflection, which in turn implies that it possesses wavelike features. Burst-like character of unidirectional fluxes is a property specific to the dark matter fluxes subjected to gravitational focusing. The fact that the bursts were observed for radiation emitted from β sources serves as another indication that the active agent is neutrino.

### Discussion

The foregoing analysis of properties of slow cosmic neutrinos was conducted without introduction of any "new entities", but rather by combining elements of celestial mechanics and quantum physics with results of astronomical observations. This analysis yielded some testable conclusions that were confirmed experimentally (periodic variations of β radioactivity; count rate bursts for a β radioactive source mounted at the focal point of a parabolic mirror; influence exerted by diffraction gratings). Based on the experimental results, an estimate of the neutrino mass was obtained.

Note that this estimate (about 20 $eV$) does not agree with published results of measurements of the mass of the electron neutrinos emitted in tritium decays (less than 2.2 $eV$) [19,23]. An important feature of this experiment warrants a special consideration. The measured spectrum of β particles near the maximum energy differs in a significant way from the theoretical one. Moreover, results change with every new measurement with annual rhythm (fig. 5, top). The upper bound of 2.2 $eV$ for the neutrino mass was obtained after introduction of corrections with the aim to make the measured spectrum more consistent with the theoretical one.

In order to explain this unexpected result, a hypothesis was put forward in [19], that spontaneous β decay of tritium is superimposed by reverse β decay, induced by a neutrino cloud around the Sun. «The proximity of the oscillation period of the step (bump) to period of Earth circulation around the Sun and other features of the phenomenon allows one to remind speculation about an effect produced by capture of the cosmological neutrino by tritium atoms with emission of almost monochromatic electrons… The Earth in its movement produces the periodical modulation of binding energy and accordingly position of the step». So, results of the tritium experiments lead to the ideas resembling the ones that were expounded by us above.

Note that averaged curves of count rates of β sources in our experiments are similar to the curves of magnitude variations of the effect discovered in the Troitsk experiment (fig. 5). This serves as another demonstration of close affinity between results obtained during our long-term measurements of radioactivity and measurements of the tritium spectrum.

The bound of 2.2 *eV* for the neutrino mass arises due to ignoring the peak near the end of the spectrum, which is presumably caused by absorption of cosmic neutrinos. At the same time, to explain such strong oscillations (up to 20 *eV*) in the results of unfiltered measurements, one has to assume that the absorbed particle has a mass in excess of that value. Thus, a rather strange picture emerges: emitted antineutrino has a mass less than 2.2 *eV*, but absorbed neutrino has a mass over 20 *eV*. This contradiction should be resolved in the future. Note that the mass estimate for the neutrino absorbed by tritium agrees with our figure of about 20 *eV* obtained by analysis of annual rhythms of radioactivity and experiments with diffraction gratings.

It should be noted that neutrinos do not necessarily possess a single mass value. There is a suggestion that the electron neutrino may have several different mass values up to 17 *keV* and some authors even speculate that there exist a number of weakly interacting particles with different masses [20-23]. Current theoretical and experimental situation leaves enough room to accommodate various estimates of the neutrino mass.

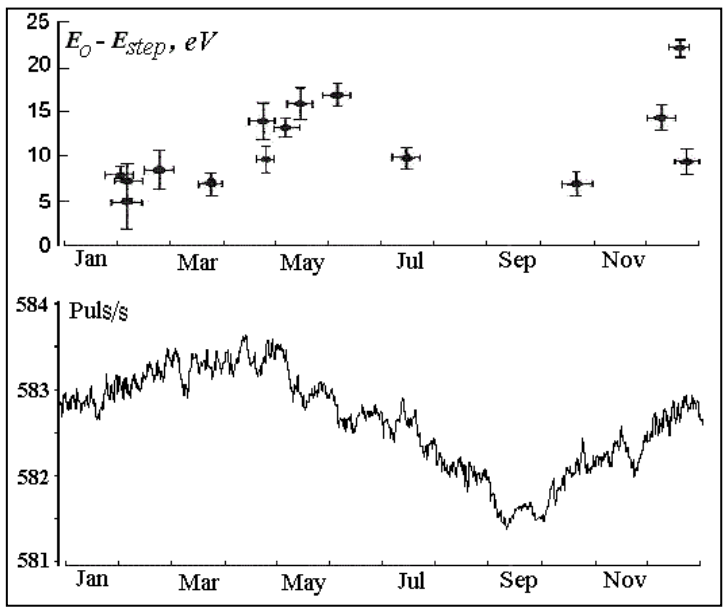

**Fig. 5.** Top: the Troitsk experiment. Difference between theoretical and experimental ranges of the β decay spectrum of tritium near maximum energy as a function of the time of year (1994-2001) [19]. Bottom: averaged count rate of a β source $^{90}$Sr-$^{90}$Y (2000-2007) [3].